\documentclass[12pt]{article}
\usepackage{amsmath2000}
\allowdisplaybreaks
\def\beq{\begin{equation}}
\def\eeq{\end{equation}}
\def\bea{\begin{eqnarray}}
\def\eea{\end{eqnarray}}
\def\del{\partial}

\def\b{\beta}
\def\D{\Delta}
\def\l{\lambda}

\def\a{\alpha}

\def\b{\beta}
\def\D{\Delta}
\def\l{\lambda}

\def\g{\gamma}

\def\ket{\rangle}
\def\D{\Delta}
\def\L{\Lambda}
\def\href#1#2{#2}

\def\3{\sigma_3}
\def\1{\sigma_1}

\def\be{\begin{equation}}
\def\ee{\end{equation}}
\def\ba{\begin{eqnarray}}
\def\ea{\end{eqnarray}}

\def\Tr{{\rm Tr}}
\newcommand{\bra}[1]{\langle #1\vert}
\newcommand{\calO}{{\cal O}}
\newcommand{\calN}{{\cal N}}

\def\del{\partial}

\def\b{\beta}
\def\D{\Delta}
\def\l{\lambda}

\def\a{\alpha}

\def\b{\beta}
\def\D{\Delta}
\def\l{\lambda}

\def\g{\gamma}

\def\ket{\rangle}
\def\L{\Lambda}
\def\O{{\mathcal O}}

\begin{document}

\pagestyle{plain} \setcounter{page}{1}
\begin{titlepage}

 \rightline{\small{\tt CALT-68-2390}}
 \rightline{\small{\tt CITUSC/02-021}}
 \rightline{\tt hep-th/0206065}

\vskip -1.0cm


\begin{center}

\vskip 2 cm

{\LARGE {Cubic Interactions in PP-Wave \\
\vskip .2cm Light Cone String Field Theory}}

\vskip 2cm {\large Peter Lee, Sanefumi Moriyama and Jongwon Park}

\vskip 1.2cm

\end{center}

\begin{center}
\emph{ California Institute of Technology 452-48, Pasadena, CA
91125}

\vskip .7cm {\tt peter, moriyama, jongwon@theory.caltech.edu}

\end{center}
\vspace*{1in}
\begin{center}
\textbf{Abstract}
\end{center}
\begin{quotation}
\noindent We use the supergravity modes to clarify the role of the
prefactor in the light-cone superstring field theory on PP-wave
background.  We verify some of the proposals of the recent paper
{\tt hep-th/0205089} and give further evidence for the
correspondence between ${\cal N}=4$ SYM gauge theory and string
theory on PP-wave. We also consider energy-preserving processes
and find that they give vanishing cubic interaction Hamiltonian
matrix.

\end{quotation}
\vfil
\end{titlepage}

\newpage


\section{Introduction}
Recently, Berenstein, Maldacena, and Nastase \cite{BMN} came up
with a very exciting proposal where we can test AdS/CFT
correspondence beyond the supergravity approximation. It is based
on the discovery \cite{M,MT} that Green-Schwarz strings on pp-wave
background are exactly solvable in the light-cone gauge, and the
observation \cite{BFHP1,BFHP2} that the pp-wave background can be
obtained from $AdS_5 \times S^5$  in the Penrose limit. Via the
AdS/CFT dictionary, the authors of \cite{BMN} have identified the
corresponding limit in ${\mathcal N}=4$ super Yang-Mills and
argued that Type IIB sting theory on the pp-wave background is
dual to a sector of operators with large R-charge $J\sim\sqrt{N}$
and finite $\Delta-J$ in the limit $N\rightarrow \infty$ while
keeping $g_{YM}$ fixed. In this limit, although the usual 't Hooft
coupling $g^2_{YM}N$ goes to infinity, perturbative SYM is
well-defined due to the near BPS property of operators under
consideration. In particular, the duality allows one to compute
the {\it free} string spectrum from perturbative SYM calculation
\cite{BMN,GMR,KPSS,largeN}.

It is a very important and fascinating question whether this
success can be extended to the {\it interacting} string theory.
However, the holographic idea of AdS/CFT \cite{Maldacena,GKP,W}
does not seem to be directly applicable here and the first
principle is not yet available.\footnote{For some recent progress
in this direction, see \cite{DGR,KP,LOR,BN}.} Without fully
understanding it, the natural frame work for the interacting
string theory on pp-wave is believed to be the light-cone string
field theory, and the authors of \cite{SV} have constructed the
cubic interaction Hamiltonian following the light-cone string
field theory formalism of \cite{GS, GSB}. The cubic interaction
Hamiltonian is roughly speaking a three-string delta functional
with a prefactor which is argued to be the same as that in flat
spacetime.

Shortly after this development, the corresponding SYM objects are
proposed to leading order in $g^2_{YM}N/J^2=1/(\mu\a' p^+)^2$ to
be \cite{largeN}
 \beq
  \label{pro}
  \langle 1\,2\,3 |H_3 \rangle = \mu (\Delta_3-\Delta_1-\Delta_2)
  C_{123},
 \eeq
where $\Delta_i$ is the conformal dimension of the corresponding
gauge theory operator ${\mathcal O}_i$ and $C_{123}$ is the
coefficient of the three point function in planar and free theory
limit.  It should be emphasized that the proposal is limited to
processes where light-cone energy is preserved to leading order in
$g^2_{YM}N/J^2$ so that it can be captured in the perturbative
SYM\footnote{ It is argued in \cite{largeN} that the process where
light-cone energy is not preserved in the leading order in
$g^2_{YM}N/J^2$ corresponds to non-perturbative effects in gauge
theory side.}.  In this article, we perform an explicit check of
this proposal for supergravity modes. Note that for supergravity
modes, if light-cone energy is preserved to leading order in
$g^2_{YM}N/J^2 = 1/(\mu\a' p^+)^2$, it is preserved exactly since
there are no corrections.  In \cite{largeN}, it is further
conjectured $C_{123}$ corresponds to the three-string interaction
vertex and the dressing factor $\mu (\Delta_3-\Delta_1-\Delta_2)$
is reproduced by the prefactor. We explicitly check that this is
the case for particular bosonic excitations considered in
\cite{largeN} and extend the proposal to other excitations.

This paper is organized as follows. In section 2, we briefly
review relevant materials and fix our convention. In section 3, we
discuss the prefactor and in section 4, we compute bosonic
three-string Hamiltonian matrix elements and compare them to
three-point functions on the gauge theory side. In section 5, we
consider general cubic interactions and in section 6 we end with
discussion.

While this manuscript was being prepared, some related articles
\cite{KKLP, H, CKT} have appeared on the archive. The authors of
\cite{KKLP} fix the normalization of the cubic Hamiltonian
constructed by \cite{SV}, and then compute the matrix elements of
chiral primaries and find agreement with SYM calculations as in
(\ref{pro}). Some parts in section 4 has been first computed in
\cite{H} by dropping the prefactor. Our work clarifies this point
and considers general supergravity matrix elements.

\section{Review}
\subsection{PP-wave string / SYM correspondence}
In this section, we briefly review pp-wave/SYM correspondence and
fix our notation and convention. The pp-wave background is
obtained by taking the Penrose limit of  $AdS_5 \times S^5$ and is
given as \beq ds^2 = -4dx^+dx^- -\mu^2\left( \sum_{i=1}^8
x^ix^i\right) dx^{+2}
 +\sum_{i=1}^8 dx^{i\,2}
\eeq with Ramond-Ramond flux \beq F_{+1234} =F_{+5678}=2\mu. \eeq
The metric has $SO(8)$ rotational symmetry of $x^i$'s but it is
explicitly broken to $SO(4)\times SO(4)$ by the RR flux. The
Green-Schwarz string in this background is exactly solvable in the
light-cone gauge \cite{M,MT} and the spectrum is a tower of free
massive harmonic oscillators : \beq H_2 = {1\over \a^\prime p^+}
\sum_{n=-\infty}^{\infty} \omega_n\left( \sum_{i=1}^8
a_n^{i\,\dagger} a_n^i + \sum_{a=1}^8 b_n^{a\,\dagger}
b_n^a\right), \eeq where $\omega_n = \sqrt{n^2+(\a^\prime\mu
p^+)^2}$.  BMN propose that the light-cone vacuum state is dual to
a chiral primary operator
 \begin{align}
   |{\rm vac}\rangle \quad&\longleftrightarrow \quad
   \frac{1}{\sqrt{J}\sqrt{N^J}} \Tr \left[ Z^J\right] \,.
 \end{align}
For zero modes or supergravity modes, we insert proper operators
with $\Delta-J=1$ at all possible positions in the vacuum
operator. For excitations along 1,2,3,4 direction we insert $D_i
Z$, and for 5,6,7,8 directions, $\phi^i$.
\subsection{ PP-wave light-cone string field theory}
The authors of \cite{SV} construct the cubic interaction
Hamiltonian $H_3$ following the light-cone string field theory
formalism of \cite{GSB}. It can be expressed as \beq
 |H_3\rangle= \hat{h}_3|V\rangle\,.
\eeq $|V\rangle$ is just a kinematical three string delta
functional which preserves kinematical symmetries and is common to
the other dynamical generators. $\hat{h}_3$ is a prefactor
inserted at the interaction point to respect the whole
supersymmetry algebra including dynamical symmetries. In
\cite{SV}, the prefactor is claimed to be of the same form as in
flat spacetime since the prefactor arises from a worldsheet UV
effect and the additional mass term in the pp-wave should not
affect it. More explicitly,  they are given as
 \beq
   |V\ket = E_a E_b |0\ket\,, \hskip 1cm   \hat{h}_3 = {\mathsf P}^i {\mathsf P}^j v_{ij}(\L)
 \eeq
Let us define $\a_r\equiv \a^\prime p^+_r$. With $\a=\a_1 \a_2
\a_3$ and $\b=\a_1 / \a_3$, we have
 \begin{align}
  & E_a = \exp\left[{1\over 2} \sum_{r,s=1}^3 \sum_{i=1}^8 a_r^{\dagger i}M^{rs}a_s^{\dagger i}\right],
\; M=\begin{pmatrix}
  \b+1 & -\sqrt{-\b(1+\b)} & -\sqrt{-\b} \\
  -\sqrt{-\b(1+\b)}  & -\b & -\sqrt{1+\b} \\
  -\sqrt{-\b} & -\sqrt{1+\b} & 0
\end{pmatrix}, \\ \nonumber
  & E_b = \l^1...\l^8, \hspace{1cm} \l=\l_1+\l_2+\l_3, \hspace{1cm}
  {\mathsf P}^i=\a_1 p^i_2 - \a_2 p^i_1, \hspace{1cm} \L=\a_1 \l_2 -
\a_2 \l_1,
   \\
  & v_{ij}(\L)=\delta^{ij}+{1\over 6 \a^2}\g^{ik}_{ab} \g^{jk}_{cd} \L^a\L^b\L^c
  \L^d+{16\over 8! \a^4}\delta^{ij}\epsilon_{abcdefgh}\L^a\L^b\L^c\L^d\L^e\L^f\L^g\L^h.
 \end{align}
Here, $i,j$ and $a,b$ are $SO(8)$ vector and spinor indicies
respectively\footnote{Here the reader is not to be confused with
$E_a, E_b$ where indices $a,b$ refer to bosonic and fermionic part
of the prefactor respectively.}. We take $\a_1,\a_2>0$ and
$\a_3<0$ such that $\a_1+\a_2+\a_3=0$. In addition,
$\gamma^{ik}_{ab}$ are the usual anti-symmetrization of gamma
matrices and for concreteness we use a basis such that \beq
\g_1\g_2\g_3\g_4=\begin{pmatrix}
  1_4 & 0 \\
0& -1_4
\end{pmatrix}.
 \eeq
In this basis, $\l$ takes the following form in terms of harmonic
oscillator operators
 \bea
   && \l_r = \sqrt{\a_r\over 2}\Big(
 b_r^{\dagger 1}\;b_r^{\dagger 2}\;b_r^{\dagger 3}\;
  b_r^{\dagger 4}\;
  b_r^5\; b_r^6\; b_r^7\;b_r^8 \Big)^T, \;\;\;r=1,2 \\
  && \l_3 = \sqrt{-\a_3\over 2}\Big(
  b_3^1\; b_3^2\; b_3^3\; b_3^4\;
  b_3^{\dagger 5}\; b_3^{\dagger 6}\; b_3^{\dagger 7}\;
  b_3^{\dagger 8}
  \Big)^T.
 \eea
Lastly the "ground" state $|0\ket$ is related to the "vacuum"
state as \cite{MT}
 \beq
  |0\ket=b_1^{\dag 5} b_1^{\dag 6}b_1^{\dag 7}b_1^{\dag 8}
  b_2^{\dag 5}b_2^{\dag 6} b_2^{\dag 7} b_2^{\dag 8} b_3^{\dag 1}
  b_3^{\dag 2} b_3^{\dag 3} b_3^{\dag 4} |{\rm vac}\ket.
 \eeq
\subsection{$H_3$ from perturbative SYM}
Let $|1\rangle, |2\rangle,|3\rangle$ be free single string states
with unit norm and $\O_1,\O_2,\O_3$ be the corresponding gauge
operators with unit two-point function : \beq \langle
\bar{\O}_i(0) \O_j(x) \rangle = \frac{\delta_{ij}}{(2\pi
x)^{2\Delta_i}}\,. \eeq Define $C_{123}$ as \beq
 \langle \O_1(x_1)\O_2(x_2){\bar\O}_3(x_3) \rangle = \frac{\delta_{J_3,J_1+J_2} C_{123} }{ (2\pi x_{13})^{\Delta_3 +\Delta_1-\Delta_2}  (2\pi x_{23})^{\Delta_3 +\Delta_2-\Delta_1}  (2\pi x_{12})^{\Delta_1 +\Delta_2-\Delta_3} }\,,
\eeq in the planar limit. In the regime of small $g^2_{YM}N/J^2$,
the authors of  \cite{largeN} propose to leading order in
 $g^2_{YM}N/J^2$ that
\beq
  \langle 1\,2\,3 |H_3 \rangle = \mu (\Delta_3-\Delta_1-\Delta_2) C_{123}\,.
\eeq This proposal has successfully reproduced mass
renormalization of excited string states via second-order quantum
mechanical perturbation theory and thereby passed unitarity check.
It is further conjectured that the prefactor reproduces the
dressing factor while $|V\rangle$ corresponds to $C_{123}$. One
should note that the proposal has been applied for matrix elements
of bosonic string excitations along directions 5 to 8 only. The
authors of \cite{largeN} have claimed that for general bosonic
excitations, the prefactor gives a factor of the form
 \beq
  \mbox{(net \# of insertions along directions 1-4) } minus \mbox{ (net \# of
  insertions along directions 5-8)}.
 \eeq
By "net \#", we mean ( \# in operator 1 $+$ \# in operator 2 $-$
\# in operator 3 ). Lastly, they have argued that the prefactor
should lead to a vanishing result for fermionic excitations only.\\
In the following sections, we confirm this set of proposals for
the matrix elements of the cubic interaction Hamiltonian, $H_3$,
restricted to the supergravity sector.

\section{Prefactor}
In this section, we illuminate the role of the prefactor term in
the three-string interaction Hamiltonian.  Namely, we validate the
conjecture of \cite{largeN} by showing that the prefactor
factorizes as expected in the literature only when one considers
bosonic excitations without fermionic ones.  In the appendix, we
show that the expression for the prefactor simplifies due to the
following relation
 \beq
 \label{prefactor}
    {\mathsf P}^i{\mathsf P}^j E_a |{\rm vac}\ket=\mu\a_1\a_2\left(\a_2
    a_1^{\dagger i} a_1^{\dagger j} + \a_1 a_2^{\dagger i}
    a_2^{\dagger j} - \sqrt{\a_1\a_2} \left(a_1^{\dagger i}a_2^{\dagger
    j}+a_1^{\dagger j}a_2^{\dagger i}  \right) \right) E_a |{\rm vac}\ket.
 \eeq

Let us first consider purely bosonic excitations.  In this case,
$v^{ij}$ in the prefactor simplifies to \cite{GS1}
 \beq
  \label{prefac3}
  v^{ij}={1\over 6 \a^2}\g_{ab}^{ik}\g_{cd}^{jk}\L^a\L^b\L^c\L^d.
 \eeq
and only the $t^{ij}_{5678}\L^5 \L^6 \L^7 \L^8$ term survives in
the prefactor, where
 \beq
  t^{ij}_{5678}\equiv \g^{ik}_{\left[56\right.}
  \g^{jk}_{\left.78\right]}=\left(
  \begin{matrix}
  -1_4 & 0 \\ 0& 1_4 \\
  \end{matrix} \right)
 \eeq
in the basis of gamma matrices we are using.  Hence, for purely
bosonic amplitudes, we only need to focus on the case when $i=j$.
By employing a similar strategy as outlined in the appendix,
relation (\ref{prefactor}) with $i=j$ can be written
as\footnote{One can also see this relation for the case $i=j$ by
using the fact that $\left(p_1+p_2+p_3\right)|V\ket=0$ and
$\a_1+\a_2+\a_3=0$. We thank M. Spradlin and A. Volovich for
pointing this out.}
  \beq
 \label{bosonic}
    {\mathsf P}^i{\mathsf P}^i E_a |{\rm vac}\ket=-\mu\a\left(
    a_1^{\dagger i} a_1^{i} + a_2^{\dagger i}
    a_2^{i} - a_3^{\dagger i}a_3^i \right) E_a |{\rm vac}\ket.
 \eeq
Therefore, the prefactor takes the following form
 \beq
  \label{bosonicprefactor}
  {\mu\over 6 \a}\left[\sum_{i=1}^4\left(a_1^{\dagger i} a_1^{i}
   +a_2^{\dagger i} a_2^{i}-a_3^{\dagger i} a_3^{i} \right)
   -\sum_{i=5}^8\left(a_1^{\dagger i} a_1^{i}
   +a_2^{\dagger i} a_2^{i}-a_3^{\dagger i} a_3^{i} \right)\right].
 \eeq
The light-cone Hamiltonian is given as $\mu\sum_i a^{\dagger i}
a^i$, and we have shown that for only bosonic zero-mode
excitations along directions 5 to 8, the prefactor becomes
 \beq
  \hat h_3\sim\left(p_3^--p_1^- -p_2^- \right)
 \eeq
upto an overall constant factor.  Recall that bosonic excitations
along directions 5 to 8 correspond to insertion of scalar defects
on the SYM side.  Furthermore, for generic bosonic overlaps, the
prefactor is
 \beq
  \label{delta}
   \hat h_3\sim \mu \left( (\hat\D_1+\hat\D_2-\hat\D_3)
     - (\tilde\D_1+\tilde\D_2-\tilde\D_3)\right),
 \eeq
where $\hat\D$ stands for number of bosonic excitations along
direction 1 to 4 and $\tilde\D$ stands for the number along
directions 5 to 8.

Next we consider purely fermionic excitations.  One can
immediately see that three-string Hamiltonian matrix elements
vanish due to the creation operator coming from (\ref{prefactor})
which acts to the left on the three-string {\rm vac}uum.  Since
${\mathsf P}^i{\mathsf P}^j$ vanishes for all $i,j$, the three
string fermionic amplitude vanishes.

One should note that the factorization of the prefactor term
occurs here because there is no fermionic excitation.  Once we
include fermionic modes such that $a,b,c,d$ indices take value in
both SO(4) subgroups of SO(8), $t^{ij}_{abcd}$ matrices become
nondiagonal.  In such cases, the prefactor in general may not
factorize in a simple form given in (\ref{delta}).

\section{Supergravity vertex and SYM three-point function}
Having clarified the effect of the prefactor in the previous
section, let us proceed and compute the on-shell cubic interaction
Hamiltonian matrix elements for bosonic supergravity modes.  We
restrict to the excitations along direction 5 to 8 in order to
match the matrix elements with the gauge theory three-point
functions. We have explicitly shown that the prefactor factorizes
as in (\ref{delta}). Dropping this overall factor, an on-shell
process involving only the bosons can be evaluated to give \be
\label{sugraamp} \langle {\rm vac}\vert\prod_{i=5}^8
\frac{\bigl(a_1^i\bigr)^{l_i}}{\sqrt{l_i!}}
\frac{\bigl(a_2^i\bigr)^{m_i}}{\sqrt{m_i!}}
\frac{\bigl(a_3^i\bigr)^{n_i}}{\sqrt{n_i!}} E_a\vert {\rm vac}
\rangle/ \langle {\rm vac}\vert E_a\vert {\rm vac}\rangle
=\prod_{i=5}^8\delta_{l_i+m_i,n_i}\frac{n_i!}{\sqrt{l_i!m_i!n_i!}}
\bigl(M^{13}\bigr)^{l_i}\bigl(M^{23}\bigr)^{m_i} \ee with the
on-shell condition $\sum_{i=5}^8(l_i+m_i-n_i)=0$. The proof is as
follows. First of all, let us note that all the directions
decompose and we only have to calculate one of eight directions
and take their products. Since $M^{33}=0$, the terms expanded from
the exponent of $E_a$ that contract with $a_{3}^i$ involve only
$a_{1}^i$ or $a_{2}^i$. Therefore, in order to obtain a
non-vanishing value, $l_i+m_i-n_i\le 0$ should be satisfied for
all the directions $i$. Since the sum over $i$ is zero,
$l_i+m_i-n_i=0$ should hold for each direction $i$. In this way,
the LHS is decomposed with respect to each direction $i$ and the
contribution of each can be treated separately. The combinatoric
factor $n_i!$ comes from determining which $a_3^i$ to be chosen as
a partner of $a_1^i$ and $a_2^i$.

Now that we have general three-point overlaps of the bosonic
supergravity excitations in the light-cone superstring field
theory, let us move to the gauge theory side to see whether these
three-point functions can be reproduced.

It was proposed \cite{BMN} that the one and two supergravity mode
excitations in the string theory side correspond on the field
theory side to
\begin{align}
a_i^{\dag}|{\rm vac}\ket\quad&\longleftrightarrow
\quad\calO^J_0\equiv\frac{1}{\sqrt{N^{J+1}}}\Tr\phi_iZ^J,\\
a_i^{\dag}a_j^{\dag} |{\rm vac}\ket\quad&\longleftrightarrow
\quad\calO^J_{00}\equiv\frac{1}{\sqrt{N^{J+2}(J+1)}}
\sum_{l=0}^J\Tr\phi_iZ^l\phi_jZ^{J-l}.
\end{align}
Here the normalization factor is determined by normalizing the
two-point function to one.  In general,
\begin{align}
\frac{a^{\dag n}}{\sqrt{n!}}|{\rm
vac}\ket\quad&\longleftrightarrow
\quad\calO^J_{0**n}\equiv\frac{1}{\calN_{J,n}} \sum\Tr\phi^nZ^J.
\end{align}
Here the summation runs over all inequivalent operators with $n$
$\phi$'s and $J$ $Z$'s and the number of them is the combinatoric
factor of choosing $n$ out of $J+n$ sites on a circle, \be
\calN_{J,n}=\sqrt{N^{J+n}(J+n-1)!/(J!n!)}. \ee

To see the correspondence with the string side, let us compute the
planar diagram of
$\langle\calO_{0**l}^{J_1}\calO_{0**m}^{J_2}\bar\calO_{0**n}^{J}\rangle$.
$U(1)_J$ charge conservation implies that $J_1+J_2=J$ for the
correlation function to not vanish. Since the proposal of
\cite{largeN} holds only for energy-preserving processes, we
restrict to the case $l+m=n$.  In any case, the non-energy
preserving three point functions scale as $1/J$ compared to the
on-shell ones and vanish in the pp-wave limit. The only thing to
do is to count the number of ways of contracting. First of all, we
do not use the cyclic symmetry of the trace for
$\bar\calO_{0**n}^J$ and write down all the terms with the factor
$1/(J+n)$. Then, we have $(J+n)$ ways to divide the string of the
operators $\bar\calO_{0**n}^J$ into the parts to be contracted
with $\calO_{0**l}^{J_1}$ and $\calO_{0**m}^{J_2}$. Since we have
included all the operators in $\bar\calO_{0**n}^J$, we have
$(J_1+l)$ inequivalent ways to contract $\calO_{0**l}^{J_1}$ and
the contraction gives us the factor $(J_1+l-1)!/(J_1!l!)$. And we
have similar factors for $\calO_{0**m}^{J_2}$. Collecting all the
factors, we find
\begin{align}
\frac{1}{\calN_{J,n}} \frac{1}{\calN_{J_1,l}}
\frac{1}{\calN_{J_2,m}}
\frac{1}{J+n}(J+n)(J_1+l)\frac{(J_1+l-1)!}{J_1!l!}
(J_2+m)\frac{(J_2+m-1)!}{J_2!m!} N^{J+n-1},
\end{align}
in all. To compare with the result of the field theory, let us
take the limit $J,J_1,J_2\to\infty$ with $J_1/J$ and $J_2/J$
fixed. Since \be \frac{(J+n-1)!}{J!}\sim J^{n-1}, \ee in the limit
$J\to\infty$, the three point function on the field theory side is
\be \langle\calO_{0**l}^{J_1}\calO_{0**m}^{J_2}
\bar\calO_{0**n}^{J}\rangle =\sqrt{\frac{n!}{l!m!}}
\biggl(\frac{J_1}{J}\biggr)^{(l-1)/2}
\biggl(\frac{J_2}{J}\biggr)^{(m-1)/2}\frac{J_1J_2}{N\sqrt{J}}. \ee
Divided by the ground state amplitude, \be
\langle\calO^{J_1}\calO^{J_2}\bar\calO^J\rangle
=\frac{\sqrt{JJ_1J_2}}{N}, \ee one has \be
\frac{\langle\calO_{0**l}^{J_1} \calO_{0**m}^{J_2}
\bar\calO_{0**n}^J \rangle}
{\langle\calO^{J_1}\calO^{J_2}\bar\calO^J\rangle}
=\sqrt{\frac{n!}{l!m!}} \biggl(\frac{J_1}{J}\biggr)^{l/2}
\biggl(\frac{J_2}{J}\biggr)^{m/2}. \ee This is exactly what is
expected from the string theory calculation given in
(\ref{sugraamp}).

\section{On-shell cubic interaction}
In this section, we discuss general on-shell cubic interaction
Hamiltonian matrix elements.  Interestingly, we find that they all
vanish. Since we are considering on-shell interactions, $v^{ij}$
in the prefactor simplifies to a single term given in
(\ref{prefac3}). First of all, consider the bosonic part of the
matrix element, which is given as a sum over terms of the form
 \beq
   \bra{{\rm vac}} (a_1)^p (a_2)^q (a_3)^r {\mathsf P}^i {\mathsf P}^j E_a|{\rm vac}\ket,
 \eeq
where the excitations can be along any of the 8 directions.  When
$r\ge p+q$, the contractions of the annihilation operators with
the creation operators in (\ref{prefactor}) will bring down at
least one factor of $M^{33}=0$.  Since every term of the form
(\ref{prefactor}) vanishes, we conclude that the full string
amplitude, even the off-shell ones, vanishes when $r\ge p+q$
independent of the fermionic part of the interaction Hamiltonian.

Next, let us consider the fermionic part of the three-string
Hamiltonian matrix element.  It is given as a sum over terms of
the form
 \beq
  \bra{{\rm vac}}(b_1)^l(b_2)^m(b_3)^n\L^a\L^b\L^c\L^d\l^1\dots \l^8
  |0\ket.
 \eeq
Let us focus on the string 3 part. First of all, denote
 \beq
  (b_3)^n \sim (\hat b_3)^{n_1} (\tilde b_3)^{n_2},
 \eeq
where $\hat b_3$ represents a fermionic zero mode excitation along
one of the directions in $\{1,2,3,4\}$ and $\tilde b_3$ represents
one in $\{5,6,7,8\}$ and $0\le n_1,n_2 \le 4$ such that
$n_1+n_2=n$. Since $\L$ involves contributions from strings 1 and
2 only, we can write the string 3 contribution to the fermionic
amplitude as
 \beq
  (\hat b_3)^{n_1}(\tilde b_3)^{n_2} (\hat\l)^4 (\tilde\l)^4 (\hat
  b^{\dag}_3)^4 |{\rm vac}\ket
 \eeq
with $\hat\l$ and $\tilde\l$ defined as that for $b$'s. From the
expression of $\l$, it is clear that $\tilde b_3$ operators must
contract with $n_2$ of 4 $\tilde \l$'s and likewise $\hat b_3$'s
must contract with $n_1$ of 4 $\hat b^{\dag}$'s to have
non-vanishing overlap. Hence, we have
 \beq
    (\hat b_3)^{n_1}(\tilde b_3)^{n_2} (\hat\l)^4 (\tilde\l)^4 (\hat
    b^{\dag}_3)^4 |{\rm vac}\ket \sim
    (\hat\l)^4 (\tilde\l)^{4-n_2} (\hat b^{\dag}_3)^{4-n_1}
    |{\rm vac}\ket.
 \eeq
Notice that this implies that $4-n_1$ of 4 $\hat \l$'s must take
the form $\hat b_3$ in order to have non-zero contribution. This
further implies that
 \beq
    (\hat b_3)^{n_1}(\tilde b_3)^{n_2} (\hat\l)^4 (\tilde\l)^4 (\hat
    b^{\dag}_3)^4 |{\rm vac}\ket
      \sim (\hat\l)^{n_1} (\tilde\l)^{4-n_2} |{\rm vac}\ket,
  \eeq
where the $b_{1,2}^{\dag}$ parts of $\hat\l$ and $b_{1,2}$ parts
of $\tilde \l$ can contribute.  Therefore, we have
 \beq
    (\hat b_3)^{n_1}(\tilde b_3)^{n_2} (\hat\l)^4 (\tilde\l)^4 (\hat
    b^{\dag}_3)^4 |{\rm vac}\ket \sim
    (b_{1,2}^{\dag})^{n_1} (b_{1,2})^{4-n_2} |{\rm vac}\ket.
 \eeq
Denoting $\L^a\L^b\L^c\L^d \sim (b_{1,2})^j (b_{1,2}^\dag)^{4-j}$
with $0\le j\le 4$, the full fermionic contribution can be written
as a sum over $j$   of terms of the form
 \beq
  \bra{{\rm vac}} (b_{1,2})^j (b_{1,2}^\dag)^{4-j} (b_{1,2})^{l+m}
  (b_{1,2})^{4-n_2} (b_{1,2}^{\dag})^{n_1} (b_{1,2}^\dag)^8  |{\rm vac}\ket.
 \eeq
In order to have non-vanishing overlap, the number of $b_{1,2}$'s
must balance the number of $b_{1,2}^{\dag}$'s. This imposes the
following condition:
 \beq
   j+l+m+4-n_2=4-j+n_1+8.
 \eeq
Since $j$ ranges from 0 to 4, we have
 \beq
   n\le l+m \le 8+n.
 \eeq
Consider a general on-shell overlap of the form
 \beq
  \bra{{\rm vac}} (a_1)^p(b_1)^l (a_2)^q(b_2)^m (a_3)^r(b_3)^n
  |H_3\ket,
 \eeq
with $l+m+p+q=n+r$.  If $l+m < n$ one immediately sees that the
amplitude vanishes trivially. When $l+m \ge n$, then $p+q \le r$
and the amplitude again vanishes since the bosonic part vanishes.
Therefore, we conclude that interestingly all energy preserving
three-string interaction Hamiltonian elements vanish. It would be
worthwhile to explore how to generalize the proposal of
\cite{largeN} to the general case we considered in this section.

\section{Discussion}
In this article, we have clarified the role of the prefactor and
the three-string delta functional in the cubic interaction
Hamiltonian $H_3$ in the zero-mode, supergravity, sector. For
purely bosonic on-shell excitations of the light-cone vacuum, the
prefactor gives the expected dressing factor where 1,2,3,4 and
5,6,7,8 directions contribute with opposite signs. The important
point of the calculation is that the prefactor contribution can be
factored out and we need only to consider the three-string delta
functional contribution, $\langle 123|V\rangle$.  Furthermore, it
is shown to correspond to the three-point function $C_{ijk}$ in
the ${\cal N}=4$ SYM gauge theory.  Here, it is crucial that two
different combinatoric considerations on the string side and on
the gauge theory side agree in the large $J$ limit.

We have also shown that for all on-shell supergravity modes
including fermionic excitations the cubic interaction Hamiltonian
matrix element vanishes. However, for generic supergravity
excitations, the role of the prefactor is not as clear as the
purely bosonic ones since it does not factorize for general
excitations. It would be interesting to explore this case further.

\subsection*{Note Added} In this paper we have used the Neumann
coefficients of the supergravity vertex $M^{(rs)}$.  However, it
has been discussed in \cite{H} that the supergravity vertex does
not match with the zero modes of the string vertex $\bar
N^{(rs)}_{00}$ in the large $\mu\a'p^+$ limit where the
perturbative gauge theory computation is valid. In general, one
has
\begin{eqnarray}
\frac{\bar N^{(rs)}_{00}}{M^{rs}}=1+\mu\alpha
B^T\frac{1}{\Gamma_+}B \equiv{\cal R} \label{N/M}
\end{eqnarray}
for $r,s=1,2$, while $\bar N^{(rs)}_{00}/M^{rs}=1$ holds with
$r=3$ or $s=3$.\footnote{The definition of $B$ and $\Gamma_+$ can
be found in \cite{SV2}.} One can evaluate the behavior of ${\cal
R}$ using the methods discussed in \cite{KSV} and find that ${\cal
R}\to 0$ as $\mu\to\infty$ and ${\cal R}\to 1$ as $\mu\to 0$.
Hence, $\bar N^{(rs)}_{00}/M^{rs}=1$ ($r,s=1,2$) holds only when
$\mu\to 0$, while for $r=3$ or $s=3$, $\bar N^{(rs)}_{00}$ equals
$M^{rs}$ for all values of $\mu$.  The physical interpretation of
this result is clear. The supergravity vertex is constructed by
assuming that the zero modes decouple from the higher ones.
Generally, this is not true because $X^{(r)}$ ($r=1,2$) has
non-vanishing overlap between the zero modes and the higher ones.
Hence, the matching between $\bar N^{(rs)}_{00}$ and $M^{rs}$
occurs only in the flat space limit $\mu\a^\prime p^+\to 0$.

In the correspondence between the string theory side and the field
theory side of sec.~4, only the non-renormalized Neumann
coefficients $\bar N^{(rs)}_{00}$ for $r=3$ or $s=3$ matters. This
corresponds to the fact that the three-point functions of the
chiral primary operators are not renormalized \cite{LMRS, DFS}.
When we identify the prefactor in the string field theory as
(\ref{delta}) in sec.~2 and appendix A, the interpolating Neumann
coefficients $\bar N^{(rs)}_{00}$ for $r,s=1,2$ also appear.
However, our claim in the present paper does not change if one
repeats the analysis using the full string field theory
three-string Hamiltonian. Details of these results will appear in
our forthcoming paper \cite{LMP2}.

\subsection*{Acknowledgment} We would like to thank Jaume Gomis,
Yoonbai Kim, Sangmin Lee, Yuji Okawa, Hirosi Ooguri and John
Schwarz for useful discussions and comments.  This research was
supported in part by DOE grant DE-FG03-92-ER40701. S.\,M.\ is
supported in part by the JSPS Postdoctoral Fellowships for
Research Abroad (\#472).

\appendix
\section{Prefactor}
In this appendix, we evaluate ${\mathsf P}^i {\mathsf P}^j
E_a|{\rm vac}\ket$.  Let us first consider the case when $i=j$ and
define\footnote{In this appendix, repeated indices are not summed
over.}
 \beq
  {\mathsf P}^i \equiv \pi^i+\pi^{\dagger i}
 \eeq
such that
   \beq
   \label{prefac}
  ({\mathsf
  P}^i)^2 E_a |{\rm vac}\ket =\left( \pi^i \pi^i + 2\pi^{\dagger i} \pi^i + \pi^{\dagger i} \pi^{\dagger
 i}+ {\a_1\a_2\mu\over 4}\a_3 \right) E_a |{\rm vac}\ket,
 \eeq
where
  \beq
  \pi^i={1\over 2}\left(\a_1 \sqrt{\a_2\mu} a_2^i - \a_2
  \sqrt{\a_1\mu} a_1^i\right).
 \eeq
Explicitly, one has
 \beq
    (\pi^i)^2={\a_1\a_2\mu\over 4}\left(\a_1 a_2^i a_2^i - 2
  \sqrt{\a_1\a_2} a_1^i a_2^i + \a_2 a_1^i a_1^i \right).
 \eeq
By using the fact that
$\left[a,f(a^\dagger)\right]=\left({\del/\del a^\dag}\right)
f(a^\dagger)$, one can straightforwardly evaluate (\ref{prefac})
term by term.  For example,
 \beq
 \label{deriv}
  a_2^i a_2^i E_a |{\rm vac}\ket = \left(M^{22}+\sum_{rs} M^{2r} M^{2s} a_r^{\dagger i}
  a_s^{\dagger i}\right) E_a |{\rm vac}\ket
 \eeq
and one gets similar expression for other terms.  Putting all the
pieces together, we have
 \begin{align}
  & (\pi^i)^2 E_a|{\rm vac}\ket = {\a_1\a_2\mu\over 4}\left\{
   \left(\a_1 M^{22}-2 \sqrt{\a_1\a_2} M^{12} + \a_2
   M^{11}\right)+ \right. \\ \nonumber
  & \left. + \left(\a_1 M^{21}M^{21}-2\sqrt{\a_1\a_2}M^{21}M^{11}
  +\a_2M^{11}M^{11}\right) a_1^{\dagger i} a_1^{\dagger i} \right. \\ \nonumber
  & \left. +2\left(\a_1M^{21}M^{22}-2\sqrt{\a_1\a_2}M^{21}M^{21}+\a_2M^{11}M^{12} \right)
  a_1^{\dagger i} a_2^{\dagger i} \right. \\ \nonumber
  & \left. + 2\left(\a_1 M^{21}M^{23}-\sqrt{\a_1\a_2}(M^{21}M^{13}
  +M^{23}M^{11})+\a_2M^{11}M^{13}\right)a_1^{\dagger i}a_3^{\dagger i} \right. \\ \nonumber
  & \left. +\left(\a_1 M^{22} M^{22}-2\sqrt{\a_1\a_2}M^{22}M^{12} +\a_2 M^{12}M^{12} \right)
  a_2^{\dagger i} a_2^{\dagger i} \right. \\ \nonumber
  & \left. +2\left(\a_1 M^{22}M^{23} - \sqrt{\a_1\a_2}(M^{22}M^{13}+
  M^{23}M^{12})+ \a_2M^{12}M^{13} \right)a_2^{\dagger i} a_3^{\dagger i} \right. \\ \nonumber
  & \left. +\left(\a_1 M^{23}M^{23}-2\sqrt{\a_1\a_2}M^{23}M^{13}
  +\a_2 M^{13}M^{13}\right) a_3^{\dagger i} a_3^{\dagger i}
  \right\}E_a|{\rm vac}\ket.
 \end{align}
The above expression can be simplified to
 \bea
  && (\pi^i)^2 E_a |{\rm vac}\ket= {\a_1\a_2\mu\over 4} \left(-\a_3 +
  \a_2 a_1^{\dagger i}a_1^{\dagger i} +
  \a_1 a_2^{\dagger i}a_2^{\dagger i} -
  2\sqrt{\a_1\a_2} a_1^{\dagger i} a_2^{\dagger i} \right) E_a |{\rm vac}\ket.
 \eea
Also,
 \beq
  (\pi^{\dagger i})^2 |{\rm vac}\ket= {\a_1\a_2\mu\over 4} \left(
  \a_2 a_1^{\dagger i}a_1^{\dagger i} +
  \a_1 a_2^{\dagger i}a_2^{\dagger i} -
  2\sqrt{\a_1\a_2} a_1^{\dagger i} a_2^{\dagger i} \right) E_a
  |{\rm vac}\ket.
 \eeq
Note that this has the same expression as above up to a constant.
Lastly, we have
 \beq
  2\pi^{\dagger i}\pi^i|{\rm vac}\ket={\a_1\a_2\mu\over 2} \left(
  \a_2 a_1^{\dagger i}a_1^{\dagger i} +
  \a_1 a_2^{\dagger i}a_2^{\dagger i} -
  2\sqrt{\a_1\a_2} a_1^{\dagger i} a_2^{\dagger i} \right) E_a
  |{\rm vac}\ket.
 \eeq
Summing over all the contributions, we conclude
 \beq
    \label{appendixprefactor}
    ({\mathsf P}^i)^2 E_a |{\rm vac}\ket=\mu\a_1\a_2\left(\a_2
    a_1^{\dagger i} a_1^{\dagger i} + \a_1 a_2^{\dagger i}
    a_2^{\dagger i} - 2\sqrt{\a_1\a_2} a_1^{\dagger i}a_2^{\dagger
    i}\right) E_a |{\rm vac}\ket.
 \eeq
The proof for the case when $i\neq j$ is analogous to this one and
the expression given in (\ref{bosonic}) holds for it as well.

\end{document}